# Libra: An Economy driven Job Scheduling System for Clusters

Jahanzeb Sherwani[1], Nosheen Ali[1], Nausheen Lotia[1], Zahra Hayat[1], and Rajkumar Buyya[2]

[1]Dept. of Computer Science and Engineering
Lahore University of Management Sciences (LUMS)
Lahore, Pakistan
{jahanzeb,nosheen,02020111,02020189}@lums.edu.pk

[2]Grid Computing and Distributed Systems Laboratory
Dept. of Computer Science and Software Engineering
The University of Melbourne, Australia
*raj@cs.mu.oz.au*

**Abstract:** Clusters of computers have emerged as mainstream parallel and distributed platforms for high-performance, high-throughput and high-availability computing. To enable effective resource management on clusters, numerous cluster managements systems and schedulers have been designed. However, their focus has essentially been on maximizing CPU performance, but not on improving the value of utility delivered to the user and quality of services. This paper presents a new computational economy driven scheduling system called Libra, which has been designed to support allocation of resources based on the users' quality of service (QoS) requirements. It is intended to work as an add-on to the existing queuing and resource management system. The first version has been implemented as a plugin scheduler to the PBS (Portable Batch System) system. The scheduler offers market-based economy driven service for managing batch jobs on clusters by scheduling CPU time according to user utility as determined by their *budget* and *deadline* rather than system performance considerations. The Libra scheduler ensures that both these constraints are met within an O(n) run-time. The Libra scheduler has been simulated using the GridSim toolkit to carry out a detailed performance analysis. Results show that the deadline and budget based proportional resource allocation strategy improves the utility of the system and user satisfaction as compared to system-centric scheduling strategies.

## 1 Introduction

Clusters of commodity computers (PCs) have emerged as mainstream parallel and distributed platforms for high-performance, high-throughput and high-availability computing [1]. They are presented together as a single, unified resource to the end users by middleware technologies such as resource management and scheduling system. Users submit jobs in batch to a resource management system queue and a centralized scheduler decides how to prioritize and allocate resources for jobs execution. To minimise the response time, the scheduling system strategy needs to prioritize competing user jobs with varying levels of priorities and importance and allocate resources accordingly. To perform these tasks effectively, schedulers require knowledge of how users value their computations and their quality of service requirements, which varies from time to time [9]; schedulers should be able to provide a feedback signal that prevents users from submitting unbounded amounts of work [11]. Unfortunately, the current approaches (e.g., Condor [2], PBS [12], SGE [15], and LSF [16]) to batch scheduling provide limited means for users to expresses their valuation of resources and quality of service requirements, if any. Also, feedback signals provide no incentive for users to pay attention to and respond to them.

To overcome the shortcomings of traditional system-centric cluster management systems, we advocate the use of the computational economy metaphor [4][17][18][19]. Computational economy enables the regulation of supply and demand of resources. It allows the users to specify QoS parameters with jobs and also offers incentive to users for relaxing QoS requirements. This essentially means that user constraints such as deadline and budget are more important in determining the priority of a job by the scheduler, than system policies such as ordering jobs according to the basis of submission time.

The rest of this paper is organised as follows. In Section 2, we discuss related work and technologies that influence or aid the development of our Libra scheduling system. In Section 3, we present architecture of Libra and economic scheduling algorithm. In Section 4, we describe the implementation of Libra as scheduler for PBS system, the simulation of Libra using GridSim toolkit, and the development of Web interface. In Section 5, we mainly present simulation results that quantify the benefits of new deadline and budget based proportional resource allocation algorithm over traditional FIFO strategy. The final section concludes the paper and highlights future work.

## 2 Related Work and Technologies

In our background research, we came across the following related work and technologies that have either influenced or have served as a base platform during the implementation and evaluation of Libra scheduler.

### 2.1 REXEC Cluster Scheduler

REXEC [10] is a remote execution environment for a campus-wide network of workstations, which is part of the Berkeley Millennium Project. At the command line, the user can specify the maximum rate (credits per minute) that he is willing to pay for CPU time. The REXEC client selects a node that fits the user requirements and executes the application directly on it. Hence, REXEC offers a generic user interface for computational economy on clusters, not a scheduling system that would allow the submission of jobs in queues according to user-defined parameters like budget and deadline, and then manage resources accordingly. REXEC allocates resources to user jobs proportional to the user valuation irrespective their job needs, whereas Libra allocates resources proportional to the needs of jobs so that more jobs can be completed by the deadline.

### 2.2 Portable Batch System (PBS) and Maui Scheduler

PBS [12] is a flexible, POSIX compliant batch queuing and workload management system originally developed by Veridian Systems for NASA. It operates on networked, multi-platform UNIX environments, including heterogeneous clusters of workstations, supercomputers, and massively parallel systems. This project was initiated to create a flexible, extensible batch processing system to meet the unique demands of heterogeneous computing networks. The purpose of PBS is to provide additional controls over initiating or scheduling execution of batch jobs, and to allow routing of those jobs between different hosts. It also has a friendly graphical interface. The default scheduler in PBS is FIFO whose behavior is to maximize the CPU utilization. That is, it loops through the queued job list and starts any job that fits in the available resources. However, this effectively prevents large jobs from ever starting. To allow large jobs to start, this scheduler implements a "starving jobs" mechanism. This method may work for some situations, but there are certain circumstances where this course of action does not yield the desired results. At this time alternative schedulers that can be used with PBS come into play. Maui is one such advanced batch scheduler with a large feature set, well suited for high performance computing (HPC) platforms [13]. It uses aggressive scheduling policies to optimize resource utilization and minimize job response time. It simultaneously provides extensive administrative control over resources and workload allowing a high degree of configuration in the areas of job prioritization, scheduling, allocation, fairness, fairshare, and reservation policies. Maui also possesses a very advanced reservation infrastructure allowing sites to control exactly when, how, and by whom resources are used. It was because of PBS's flexibility as well as stability that we chose it as the CMS on which to build our first implementation of Libra.

### 2.3 Nimrod-G Grid Resource Broker

Nimrod-G [3] is a resource management and scheduling system for wide-area parallel and distributed computing infrastructure called the Grid. The grid enables the sharing and aggregation of geographically distributed heterogeneous resources such as computers (PCs, workstations, clusters and so on), software (for special purpose applications) and scientific instruments across the Internet and presents them as a unified integrated single resource that can be widely used. Under a grid-based model for distributed computational economy, Nimrod-G serves as a resource broker and supports deadline and budget constrained algorithms for scheduling task-farming applications on the Grid. It allows the users to lease and aggregate services of resources depending on their availability, capability, performance, cost, and users' quality of service requirements. The resource price may vary from time to time and from one user to another. At runtime, the user can even enter into bidding and negotiate for the best possible cost-effective resources from computational service providers. The objective of Libra project is to expand what had been for grid computing by Nimrod-G via implementing Libra scheduler for cluster computing.

## 2.4 GridSim Toolkit

GridSim [6] is a toolkit for modeling and simulation of Grid resources and application scheduling. It provides a comprehensive facility for simulation of different classes of heterogeneous resources, users, applications, resource brokers, and schedulers. It has facilities for the modeling and simulation of resources and network connectivity with different capabilities, configurations, and domains. It supports primitives for application composition, information services for resource discovery, and interfaces for assigning application tasks to resources and managing their execution. These features can be used to simulate resource brokers or Grid schedulers for evaluating performance of scheduling algorithms or heuristics. The GridSim toolkit has been used to create a resource broker that simulates Nimrod-G for design and evaluation of deadline and budget constrained scheduling algorithms with cost and time optimizations [5].

In GridSim based simulations, the scheduling and user entities extend the GridSim class to inherit ability for communication with other entities. In GridSim, application tasks/jobs are modeled as *Gridlet* objects that contain all the information related to the job and its execution management details such as job length in MI (Million Instructions), disk I/O operations, input and output file sizes, and the job originator. The scheduler uses GridSim's job management protocols and services to map a Gridlet to a resource and manage it throughout its lifecycle.

## 2.5 PBSWeb: A Web Interface to PBS Job Scheduler

PBSWeb [14] is an excellent interface add-on for PBS, which allows submission of jobs from a registered user from anywhere on the world-wide web, as opposed to PBS's requirement of being logged onto a linux/Unix-based PC, usually physically near the cluster. Further, it is much easier to learn to use than the traditional method, and nearly all functions can be handled through it as by the conventional PBS commands. By making minor modifications to the job entry page, we were able to add our values of estimate, budget and deadline to PBSWeb.

# 3 Resource Management and Scheduling (RMS) through Libra

Resource Management and Scheduling (RMS) is the act of distributing applications among computers to maximize their throughput. It also enables the effective and efficient utilization of the resources available. The software that performs the RMS consists of two components: a resource manager and a resource scheduler. The resource manager component is concerned with problems, such as locating and allocating computational resources, authentication, as well as tasks such as process creation and migration. The resource scheduler component is concerned with tasks such as queuing applications, as well as resource location and assignment.

The Libra RMS architecture and interaction between its components is show in Figure 1. It consists of clients (e.g., user interface), server (that acts as mediator between the user and compute nodes), and a set of nodes that provide the computing horse power. Each cluster node runs a server daemon. These daemons maintain up-to-date tables, which store information about the cluster environment in which it resides. A user interacts with the RMS environment via a client program, which could be a Web browser or a customized X-windows interface. Application can be run either in interactive or batch mode, the latter being the more commonly used. In batch mode, an application run becomes a job that is submitted to the RMS system to be processed.

To submit a batch job, a user will need to provide job details to the system via the RMS client. These details may include information such as location of the executable and input data sets, where standard output is to be placed, system type, maximum length of run, whether the job needs sequential or parallel resources, and so on. In addition, Libra allows the users to express valuation of computations and QoS requirements via the parameters such as deadline and budget. Once a job has been submitted to Libra, it uses the job details and users' QoS requirements to place, schedule, and run the job.

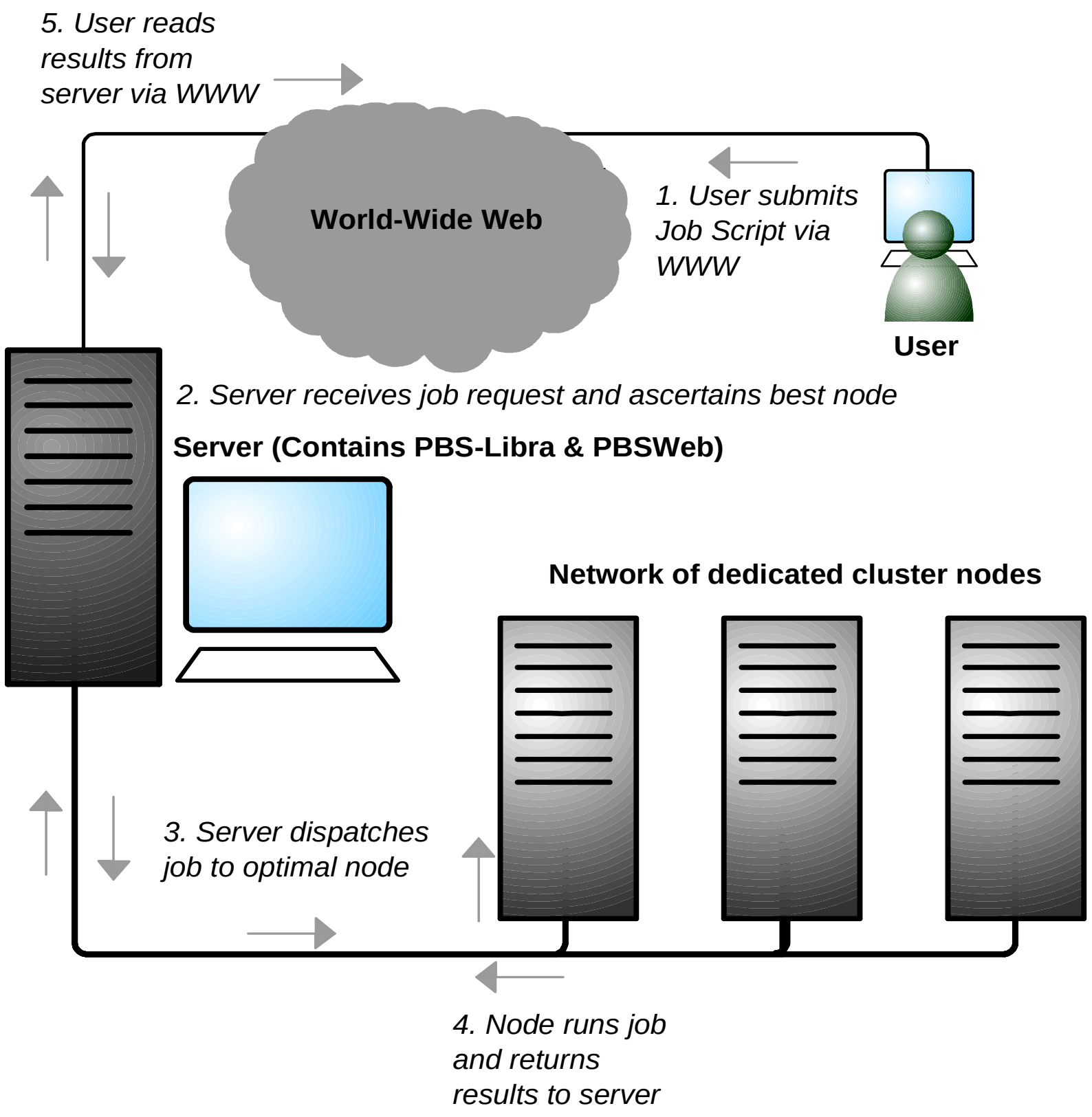

**Figure 1:** The Libra scheduler architecture.

## 3.1 Libra Scheduling Algorithm

The scheduling algorithm for the PBS-Libra Engine is described in this section. Essentially, when users submit jobs, they must submit them with values for estimated run-time (on a stand-alone node), deadline within which they require results, and budget they are willing to pay to get the job done within this deadline. These three variables will be referred to by E, D, and B, respectively. A schematic of the interaction profile between the user, CMS, scheduler, and nodes is shown in Figure 2.

The assumptions of the algorithm are as follows:

1) There is a centralized cluster management system (CMS) which is the only gateway for users to submit jobs into the cluster
2) The CMS-approved jobs are the only ones running on any cluster node (hence, nodes are dedicated and not shared between cluster jobs and other jobs)
3) When running a job, the underlying operating system accepts a parameter that is the percentage of CPU the job must be allocated, and must be able to enforce this value (as long as the sum of all shares is less than or equal to 100)
4) The estimated runtime (E) given by the user is correct for a standalone job running on any node of the cluster.
5) Homogenous cluster is assumed to have identical nodes. When nodes are heterogeneous, it is necessary to translate the user expressed E to one that represents node capability.

Thus, users submit jobs with these three values to the central server (the gateway into the cluster). There is no mechanism for users to interact with each other, and bargain on the use of resources according to their considerations, as is provided in a grid-computing environment by projects like Nimrod-G.

The server first checks whether the budget is acceptable based on a simple minimum cost formula. The formula we decided upon was:

Cost = $\alpha$*E + $\beta$*E/D (where $\alpha$ and $\beta$ are coefficients)

The value of $\alpha$ has impact on the resource pricing/job processing cost, which can be driven by supply- and demand for resources. The value of $\beta$ has impact on the incentive offered to the user for specifying the actual deadline. It offers higher incentive for users for relaxing the deadline.

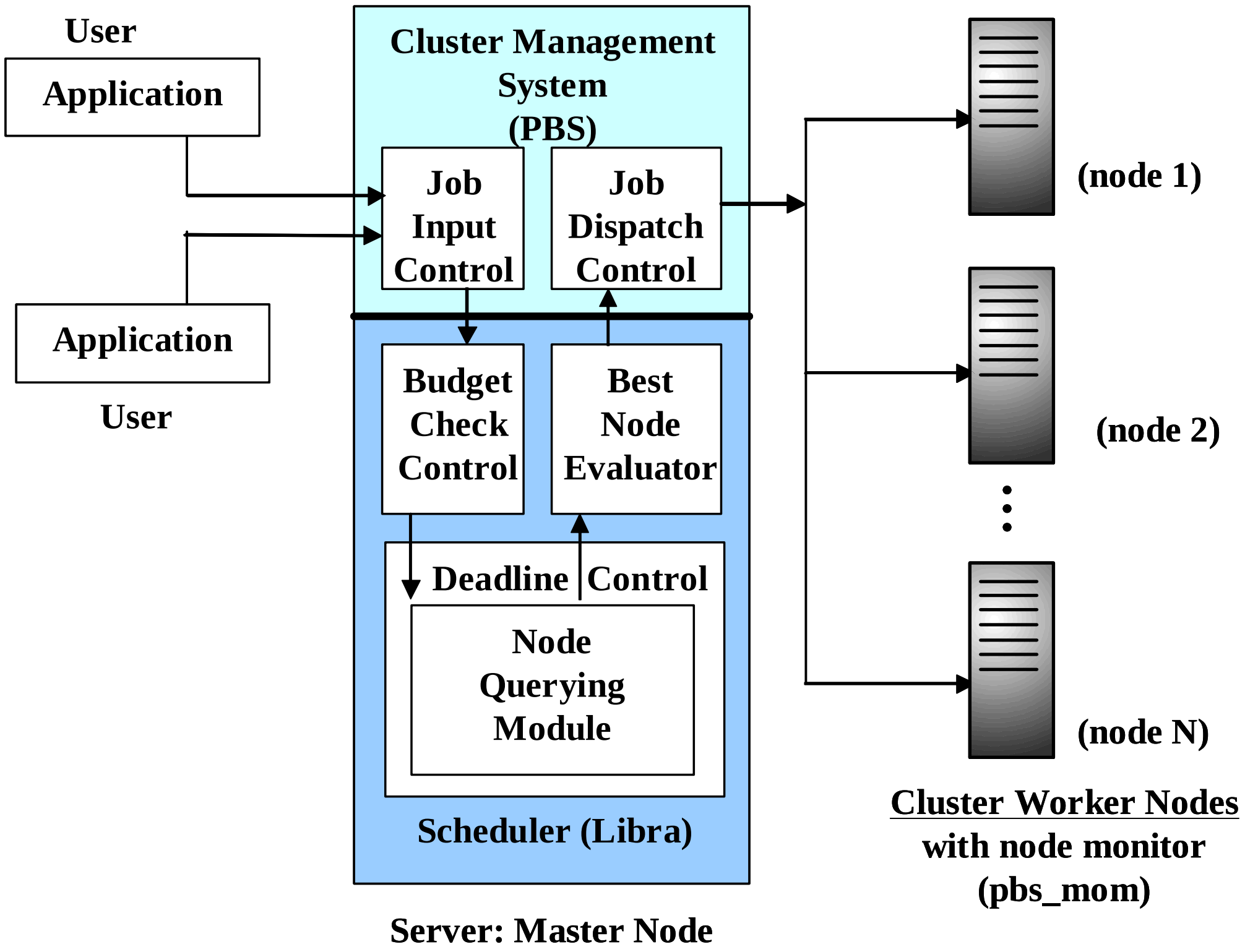

**Figure 2:** Schematic of the interaction profile between PBS and the Libra scheduler.

The logic of this cost function is as follows. For a specific job, the user must be charged on the amount of cluster hours he is using regardless of deadline – thus, for longer jobs, he will be charged more than for shorter ones. This is managed by the first term in the equation. The second term is a measure of the user's 'niceness' – the relative sacrifice he is giving in terms of the ratio of estimate to deadline. For a constant estimate, if the user increases his deadline, he is allowing us more time to handle his request, and hence, should be charged less than if he were giving us less time (and hence, requesting resources more urgently). The sum of both these values is the actual cost to the user. However, it must be borne in mind that this function may be altered by cluster managers as required for a specific set of users.

If the budget submitted by the user is acceptable, each of the cluster's nodes is queried to see whether they can complete this job within its deadline. The response by each node is either a rejection, or an acceptance, in which case the node also gives a measure of the current load it is currently servicing. Thus, if more than one node returns an acceptance, the least loaded node is the one which will finally be sent the job.

Each node keeps track of the status of jobs running on it; specifically, the CPU-time they have run for so far, and hence, the remaining run-time for the job, assuming the estimate is correct. Additionally, nodes also keep into account the arrival time of the job, and hence, the time left for the job's deadline to expire. With these two values (run-time left, and deadline left), the nodes can calculate the required CPU share that needs to be dedicated to the job so that it may complete its remaining CPU-hours within the remaining deadline. For instance, if a job has a remaining run-time of 2 CPU-hours, but a deadline of 4 realtime-hours, the node needs to allocate at least 50% of the CPU to the job over the next 4 hours for it to finish within the deadline.

Depending on the jobs already running on the node, if any, as well as the newly requested job, the node calculates the total requirement of the CPU by all of its jobs. If the sum of each job's CPU requirement is less than or equal to 100%, the job is acceptable, and the node informs the server of this. Also, the node sends the server the value of total CPU percentage as it just calculated (which is the required load on the node if the newly requested job is sent to this node).

If no node can run the job, the server informs the user that its deadline could not be met by any node, and hence the user must try later, or try again with a more relaxed deadline. However, if there is one or more node that can run the job, the job is sent to the least loaded node which responded with an acceptance. The node receives the job, and dispatches it, along with the CPU share requirement, to the operating system. Periodically (depending on the CMS and the OS), the node receives a value of CPU-time completed from the OS, and updates its internal book-keeping with this new value. If the job has run more or less than it was supposed to in this time, the node may request the OS to update its CPU share with a new value as calculated taking into account the new values of run-time left and deadline left.

Thus, by ensuring that each node is keeping within the CPU time requirements, the Libra scheduler is able to guarantee that jobs will be completed within their deadline; if any new job is to be dispatched to any node, it will only be allowed if, by virtue of its load requirements, its load requirements are possible to be met along with the other jobs running on the node.

# 4 Implementation

The implementation and evaluation of Libra scheduler and its algorithm is achieved by leveraging existing software infrastructure. The Libra scheduler has been implemented as an add-on package for PBS resource management system. The PBSWeb interface has been enhanced to support Libra's mechanism to allow users to express their valuation of computations and QoS requirements. The performance of the scheduling algorithm has been evaluated through simulation using the GridSim toolkit.

## 4.1 Libra with PBS

A schematic of the interaction profile between PBS and the Libra scheduler shown in Figure 2 illustrates how Libra as been implemented as an add-on package to the PBS. To allow the submission of E, D and B as input parameters with the job, the PBS's user input specification files src/include/site_job_attr_def.ht and src/include/site_job_attr_enum.ht have been edited. Into these, we added descriptions of the three new variables, along the lines described in these files' comments. After this, the qsub command began to accept values for E, D and B at the command line, using the -W option.

At this point, although the user could send these variables along with the job submitted to the server, the scheduler module was still running on its old considerations. The default PBS scheduler (called fifo, although it is much more than a simple fifo) works like this: it is summoned primarily every time a job is submitted, or completed. It begins by creating data structures which describe the entire cluster at that time; then, it queries the server for information to populate these data structures (most importantly server_info, node_info, job_info). Thus, we changed the data structure definitions to include E, D and B as part of job_info struct, and then also made the scheduler populate the updated structure with the values of these variables from the server.

The first thing to be done when checking for the viability of a job request was the budget, which we added to the schedule function as the first action after it begins considering the job to run. Based on a cost function [`req_budget(B,E,D)`], the budget is either rejected or accepted. If rejected, the user is told to increase his budget value to a higher amount or reduce his deadline. If accepted, the scheduler begins the next part of the scheduling process.

The actual task of ascertaining `best_node` for a job is now done, via the `accept_reject_all_nodes` function. This queries each cluster node (accessible through the `nodes` pointer) to see whether it can handle the new job's (`jinfo`) constraints based on its present constraints or not. Each node calculates the minimum required share of CPU demanded by each job assigned to the node. Each job requires a minimum CPU share of:

$$\frac{\texttt{jinfo->estimate - getCput(jinfo)}}{\texttt{jinfo->deadline - getWalltime(jinfo)}}$$

[`jinfo->estimate` is the estimate submitted by the user,

`jinfo->deadline` is the deadline submitted by the user,

`getCput(jinfo)` returns the CPU time currently executed by the job,

`getWalltime(jinfo)` returns the wall-clock time currently taken by the job]

For example, if a job has run 20 out of the 30 CPU seconds it required, and has done so in 20 seconds, with a user-given deadline of 50 seconds, it needs a minimum quantum of (30-20)/(50-20) = 33.33% of the CPU for it to complete exactly within the remaining deadline. Thus, the minimum share value for each job is calculated, as well as the minimum share for the new job that is being considered, and all the values aggregated. This value is subtracted from 1.0 and sent to the server, as `loadfree`, the idle CPU share that will remain if this job is accepted by the node. The server checks to see if there is at least one node with a non-negative value for `loadfree`. If there is no such node, the job is rejected, and the user appropriately informed. Otherwise, the `best_node` is set to the node with the least value of `loadfree`, and PBS dispatches the job off to this node as normal.

Once at the node, the `pbs_mom` module, which is responsible for executing jobs on the nodes, dispatches the job to the kernel to actually execute. At this point in time, we have only tested the algorithm on a linux kernel which does not provide a method of ensuring CPU shares to jobs (e.g. via stride scheduling or lottery scheduling). Thus, we were not able to implement the CPU shares which were calculated previously. Instead, we modified the scheduler to assume that only integral divisions of CPU were allowed (i.e. 100%, 50%, 33%, 25%, and so on), which could be approximated by the default linux kernel.

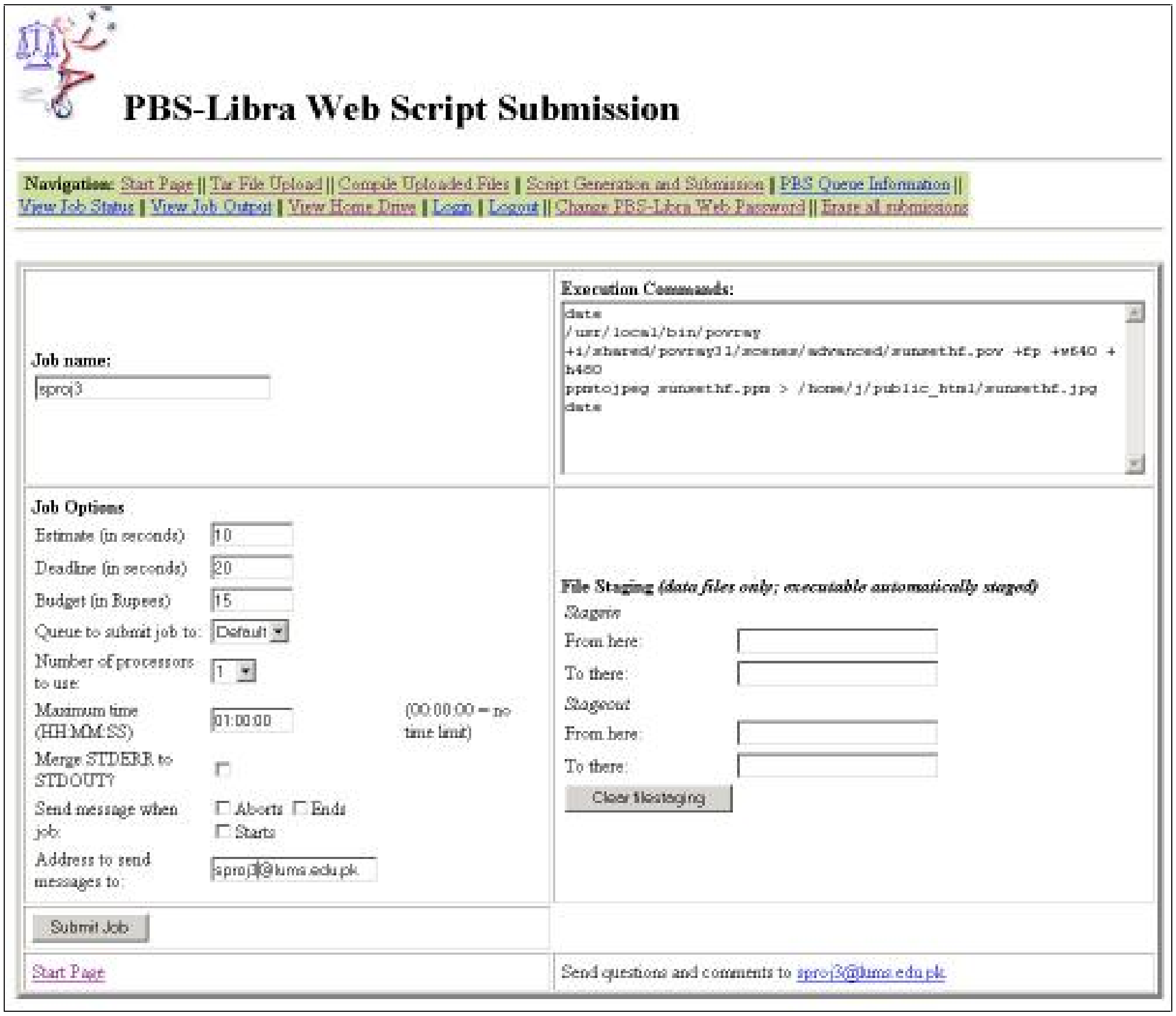

**Figure 3:** The script submission page for the PBS-Libra engine.

## 4.2 Libra with PBSWeb

For making our cluster more user-friendly and accessible, we designed an efficient front-end for the PBS-Libra engine using PBSWeb, an interface add-on to PBS developed at the University of Alberta. The PBSWeb interface was modified to suit the user-centric approach of Libra, included the provision for providing job budget, estimate and deadline with the job (see Figure 3).

The general sequence of events in running a job through PBSWeb is essentially the same as in a command-line interface: user login, job file upload, job script generation, queue status check if necessary, job status check, and job output. PBSWeb makes it easy for even novice users to use our cluster for running their jobs. Either interface can be used, depending on the user preference. To demonstrate an example, the command-line interface for the job submission command would be:

*qsub filename.pbs –w deadline=x, budget=y, estimate=z*

where the job defined by filename.pbs has been uploaded in the source directory.

PBSWeb would similarly involve the uploading of the job through the website, the specification of the job details in textboxes, and the subsequent submission of the job to the cluster through the “Submit” command. The self-descriptive snapshots of the user interface of the PBS-Libra engine provided in [20] highlight the simple and convenient use of the front-end of the PBS-Libra Engine.

## 4.3 Libra Simulation Using the GridSim Toolkit

For simulation purposes we looked to the GridSim toolkit that allows modeling and simulation of parallel and distributed system entities for evaluation of scheduling algorithms. Although, it has been developed to simulate scheduling on grids (multiple users competing for heterogeneous resources), we have been able to use it simulate scheduling on clusters with minimal effort.

First, we had to change how the resources were set up. A resource was essentially made up of a list of heterogeneous machines whose processing power was measured in terms of the MIPS rating of its processor/s. We could have simulated our homogeneous, single-processor, 20 node cluster with just one resource having a list of 20 machines each with one processor, but this would have required us to drastically change the remaining code. In order to be able to use as much of the present code as possible, we simulated each node as a separate resource. Having set up our cluster in this way in the `GridSimMain` class, we moved on to the scheduling.

GridSim uses SimJava as its base platform to create entities (each running its own thread) that are connected to and can communicate with each other. Users and resources are both entities. What the software initially did was as follows: a user would come in with an experiment (job) which was made up of many gridlets (smaller jobs that can be used when simulating embarrassingly parallel jobs), a deadline and a budget, and create his own broker who would then start looking at all the resources available to see if that user’s experiment could be completed within the given deadline and budget. The first thing that we realized was that the deadline and budget defined were *relative* values as opposed to *absolute* values; deadline was basically *how much more than the absolute minimum time required to complete the experiment* could the user afford to wait, and similarly budget was *how much more than the absolute minimum cost* could the user pay. This is not the scenario we were faced with in our work. We thus had to change those parameters both in the `Experiment` class and in the `body()` function of the `Broker` class so that the deadline and budget were the absolute values that the user submitted.

Currently, the software worked such that each user’s broker would try and schedule all of the gridlets of its user’s job on various resources so as to be able to minimize time or minimize cost, depending on which scheduling strategy was opted for. In our implementation, however, we needed *one central scheduler*. This entailed removing each user’s individual broker from the constructor of the `UserEntity` class and replacing them with one global broker that served as our cluster scheduler, and is done in the `GridSimMain` class. All the scheduling is then done in the `body()` function of the `GridBroker` class.

GridSim scheduled gridlets in batches, such that each resource would specify the optimal number that it could handle depending on its current load, and that is the number that the scheduler would schedule, and then move on to the next batch of gridlets. This allowed for some gridlets to be dispatched to resources and start executing, and then for the scheduler to discover that some of the other gridlets could not be handled on *any* of the resources within the deadline. Our implementation had to make sure that *all of the gridlets* could be completed within deadline *before* a job was accepted, and *then* dispatch them to the resource to which they had been assigned. This involved changing the order of some of the current code, allowing it to execute serially.

In addition to this, we also had to ensure that no resource was accepting more jobs (promising more MIPS) than it could handle. This meant that if two gridlets queried the same resource one after the other, and the first was accepted on that resource, then when the second gridlet examined the resource, it saw the resource's load as *including that of the previous gridlet*, even though it may not have been actually dispatched to that resource yet. This is done by defining a `reservedLoad` variable in the `BrokerResource` class that gets added to the load variable, which is the load on a resource obtained earlier. Every time a gridlet is assigned to a resource, it increments the `reservedLoad` variable by the minimum MIPS that it needs to complete within deadline:

```
Br.reservedLoad+=GridletNeedMI/Timespan;
```

In the future when the scheduler checks to see if a resource can supply the MIs that a gridlet needs within the remaining time, it accounts for the fact that a certain amount of MI have already been promised to some other gridlets in addition to its current load in the `getAvailableMIPerSec()` function in the `BrokerResource` class

```
public double GetAvailableMIPerSec()
{
  return resource.GetMIPSRatingofonePE() – LatestLoad – reservedLoad;
}
```

After the scheduler makes sure that no gridlet has been rejected by all the resources, it dispatches them to the resources to which they have been assigned and the `GridletArrives()` function in the `GridResource` class is called. This function basically updates the load on that resource and takes care of some other bookkeeping like updating how much each gridlet has completed etc...

After we had made our global broker, we realized that this broker would only move on to the next experiment after the current one had finished. To combat this, we introduced a thread in the `body()` function of the `Broker` class that would start after the gridlets had been dispatched, and allow the scheduler to move on to the next experiment.

```
eThread et=new eThread(this, experiment);
et.start();
```

Apart from this, there were minor changes made to the `Experiment` class and the `BrokerResource` class to allow the load attribute to be defined as the minimum share of the MIPS rating that was being used.

## 5 Performance Evaluation

The majority of the evaluation was carried out using GridSim, since it was impossible for us to test clusters of size greater than 4 (since we were only allotted this many by the University). The only result that was sufficiently different from what we expected was that two jobs (graphics ray-tracing jobs) running simultaneously on a node ran faster when timeshared than when run sequentially one after the other. Our assumption was that timesharing increases overhead via context switching; however, the offsetting factor could be that timesharing increases the throughput of the CPU because each job can use a different resource simultaneously (e.g., disk I/O along with math calculations).

Generating 2 batches of jobs (100 jobs, 200 jobs) of different sizes with deadlines and budgets, we ran the same data on clusters of 10 and 20 computers and compared the results obtained using two different scheduling policies: our proportional share algorithm and FIFO (First In First Out).

The $1^{st}$ batch with 100 jobs/experiments is formulated as follows. Arrival times were spread randomly over the experiments, ranging from t=1 to t=102 simulation time. Length of the jobs ranged randomly from 1000s to 10900s MI. 80% of the jobs (80) were with a budget of b=1000, the other 20 had random amounts ranging from b=1000 to b=12000. Deadlines again ranged randomly from d=1 to d=1200.

The $2^{nd}$ batch with 200 jobs/experiments is formulated as follows. Arrival times were spread randomly over the experiments, ranging from t=1 to t=208 simulation time. Length of the jobs ranged randomly from 1000s to 10900s MI. 80% of the jobs (160) were with a budget of b=1000, the other 40 had random amounts ranging from b=1000 to b=12000. Deadlines again ranged randomly from d=1 to d=1200.

The numbers of jobs accepted and rejected when they are scheduled using the default FIFO scheduling strategy of PBS and the proportional scheduling strategy of Libra is shown is shown in Table 1.

| No. of Jobs | No. of Nodes | No. of Jobs Accepted | | No. of Jobs Rejected | |
|---|---|---|---|---|---|
| | | PBS | Libra | PBS | FIFO |
| 100 | 10 | 77 | 86 | 23 | 14 |
| | 20 | 90 | 86 | 10 | 14 |
| 200 | 10 | 95 | 102 | 105 | 98 |
| | 20 | 165 | 177 | 35 | 23 |

**Table 1:** No. of jobs accepted and rejected by PBS and Libra scheduling strategies.

The results of the simulations are graphically shown in Figure 4 to Figure 7. In each graph, the blue area in the bar (i.e., the first part in the bar from the bottom up) represents the time taken for the job to complete and the red area represents the time remaining to deadline. Note that, the FIFO scheduling strategy allocates the full share of node/CPU resources to the job.

The results showed that in all four cases: small cluster with small workload (see Figure 4), small cluster with larger workload (see Figure 6), large cluster with small workload (Figure 5), and large cluster with a large workload (see Figure 7) the Libra's proportional share algorithm is performing better than the PBS's FIFO algorithm, in that it completed more jobs within the deadline. Although there isn't much of a difference between the two algorithms when the cluster is large and the workload is small, this does not mean that this is always the case. We see that as the workload increases the difference between the number of jobs accepted between the two increases.

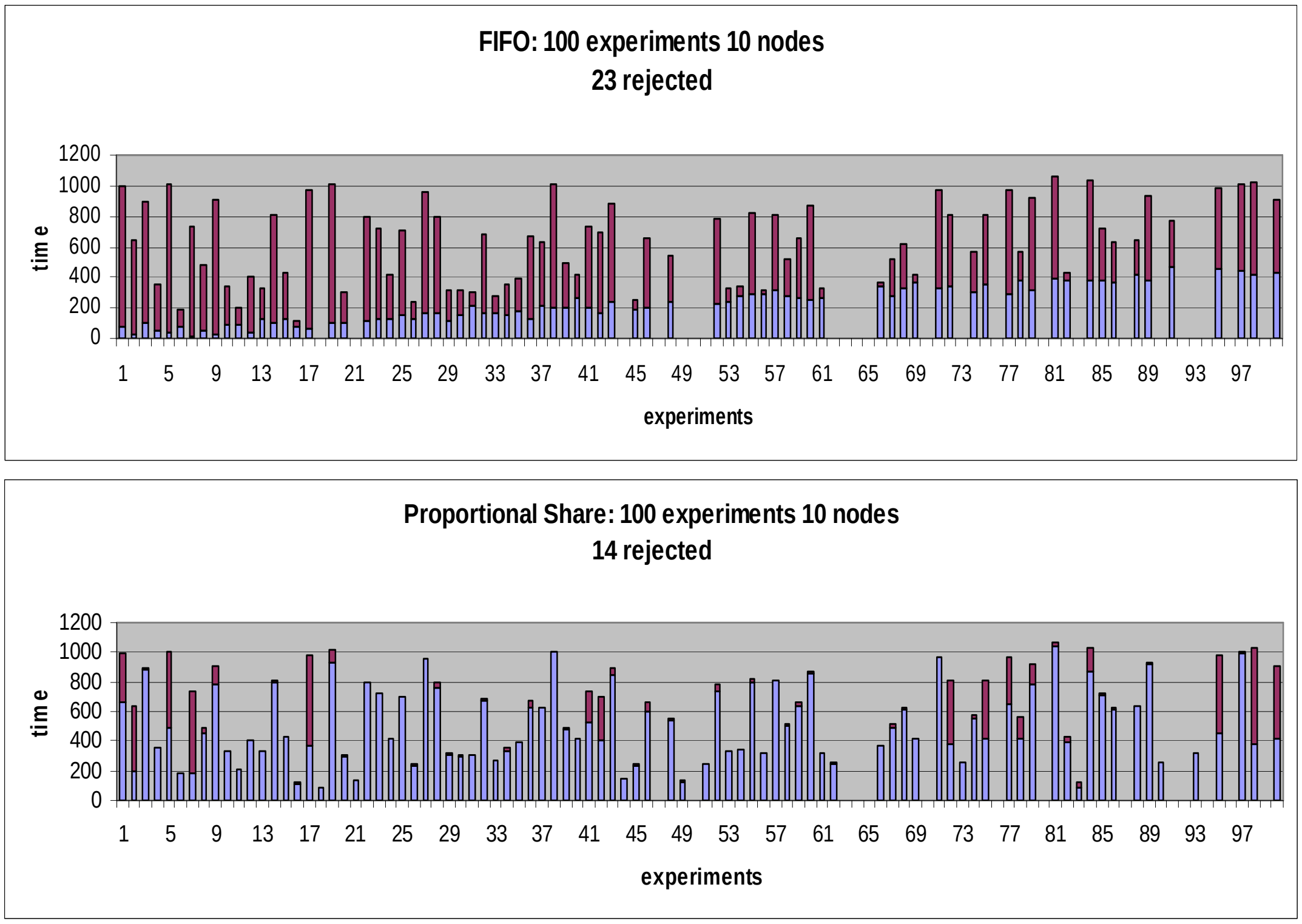

**Figure 4:** Jobs accepted and rejected by PBS and Libra scheduler: 100 jobs and 10 nodes.

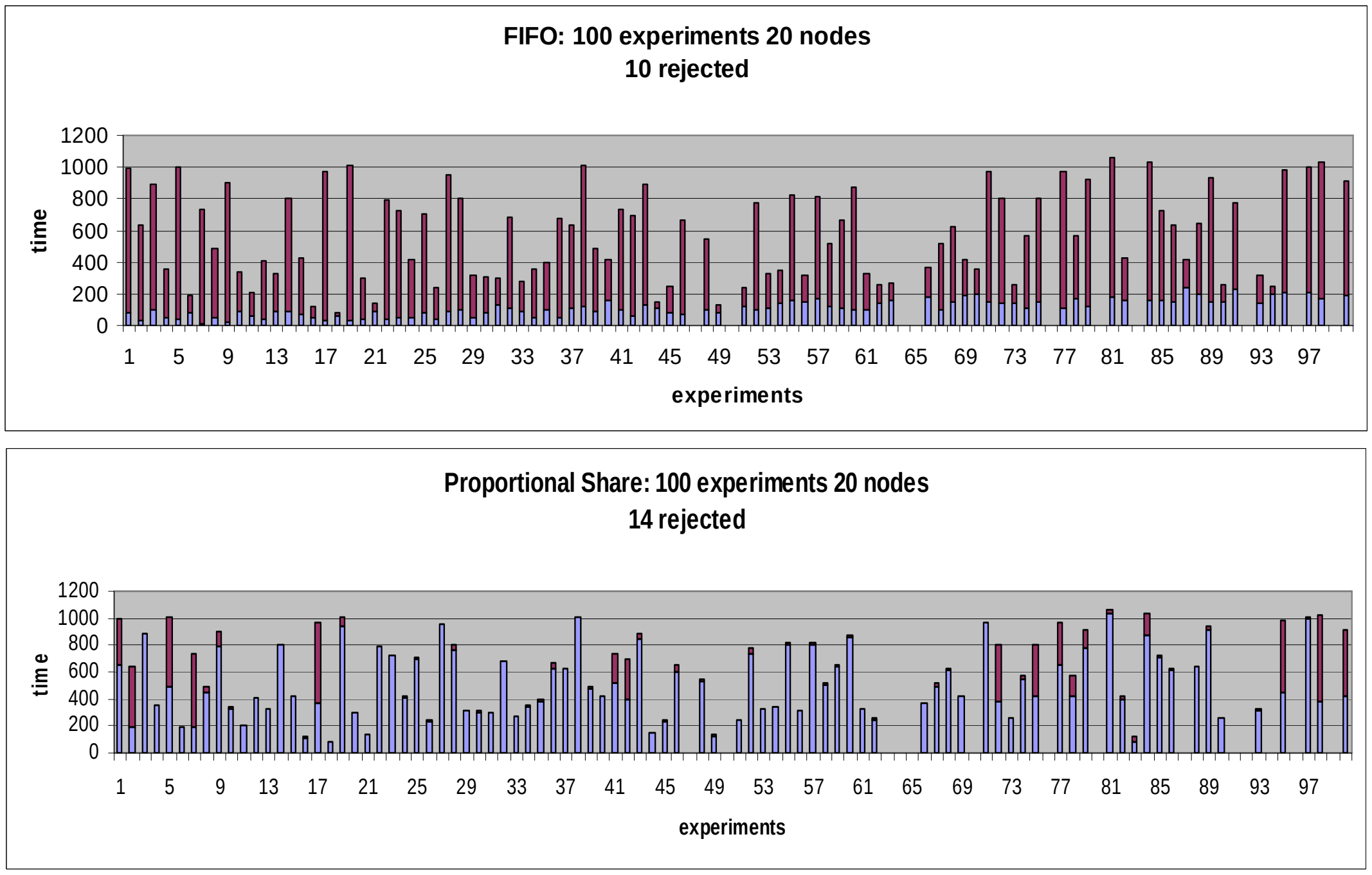

**Figure 5:** Jobs accepted and rejected by PBS and Libra scheduler: 100 jobs and 20 nodes.

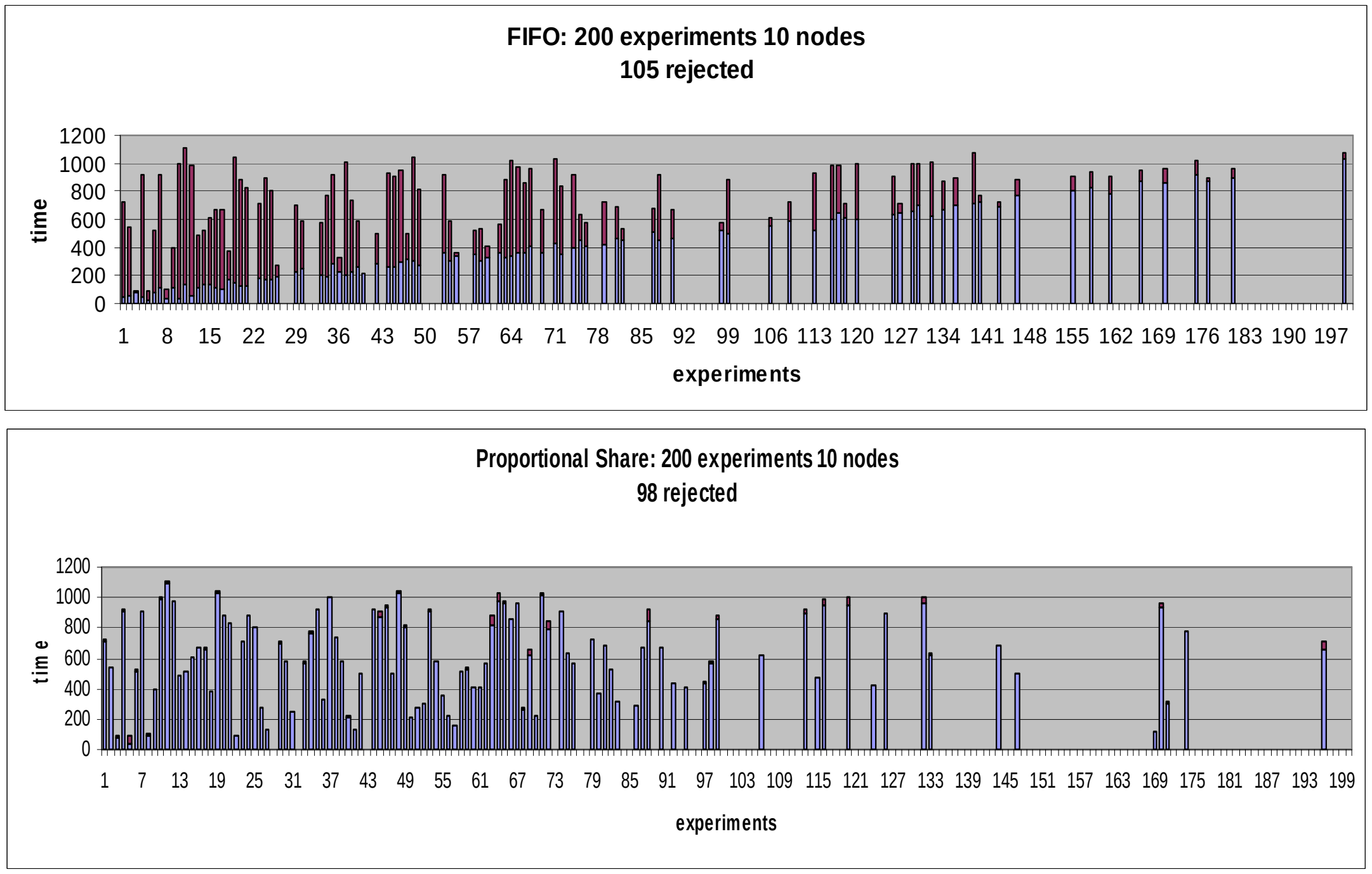

**Figure 6:** Jobs accepted and rejected by PBS and Libra scheduler: 200 jobs and 10 nodes.

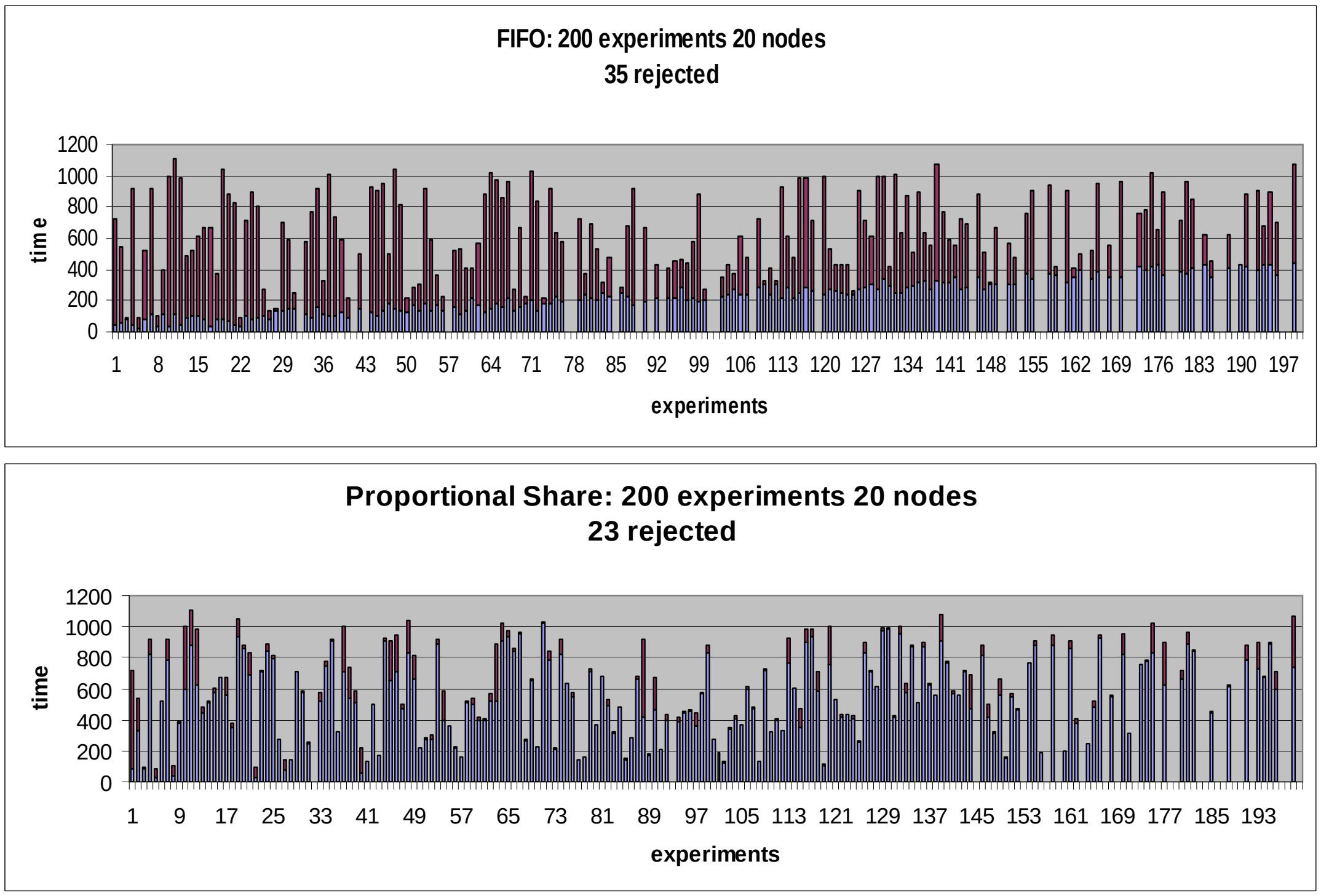

**Figure 7:** Jobs accepted and rejected by PBS and Libra scheduler: 200 jobs and 20 nodes.

## 6 Conclusion and Future Work

There is great scope for economy-based scheduling in HPC. The vast majority of users of HPC applications have little care for traditional measures such as CPU utilization; rather, the amount they pay, and the deadline they are willing to allow for the return of the job is most relevant for their purposes. Thus, architectures that incorporate such constraints would optimize user utility and hence, user satisfaction.

The algorithm we developed can be improved much further. One line of future work is that instead of time-sharing resources on one node, each node should be dedicated to each job (with more urgent jobs putting already-running jobs on hold if it leads to more jobs being handled within their deadlines, and hence, more user satisfaction), and all calculations of whether jobs can be finished within their deadlines should be carried out with that in mind. This implementation can then be checked against the current one; the factors in favor of this new algorithm include minimized context switching, since jobs are always given dedicated CPU for as long as they run, while the old one benefits from simultaneous use of different resources by timeshared jobs. In addition, the entire structure of user-based parameters needs to be checked 'in the wild', with real users submitting jobs on such a system. This would aid not only in fine-tuning the current system to users' requirements, but also in discovering user parameters not accounted for currently.

The future work investigates developing new deadline and budget driven algorithms for scheduling applications created using the task-farming model (parameter-sweep), master-worker, and parallel applications. In addition, different pricing strategies driven by supply-and-demand for resources and economic models will be investigated.

### Software Availability

The GridSim toolkit software with source code can be downloaded from the project website:

`http://www.buyya.com/libra/` or `http://www.gridbus.org`

### Acknowledgements

The lack of extensive literature on PBS's inner workings made immediately delving into actual implementation difficult. PBS's code is well-maintained and commented, however, an understanding of the "big picture" of how things worked across different modules was difficult to grasp from such comments. It was through many mails to the ever-useful PBS mailing list that we found answers to many of our (often simple) questions, and we are indebted to the people who helped us out. We thank Shoaib Burq and Srikumar Venugopal for proof reading the paper.

## References

[1] R. Buyya (editor), *High Performance Cluster Computing: Architectures and Systems*, Volume 1, Prentice Hall, USA, 1999.

[2] J. Basney and M. Livny, "Deploying a High Throughput Computing Cluster", *High Performance Cluster Computing*, R. Buyya, Editor, Vol. 1, Chapter 5, Prentice Hall PTR, May 1999.

[3] R. Buyya, D. Abramson, and J. Giddy, *Nimrod-G: An Architecture for a Resource Management and Scheduling System in a Global Computational Grid*, Proceedings of the 4th International Conference and Exhibition on High Performance Computing in Asia-Pacific Region (HPC ASIA 2000), May 14-17, 2000, Beijing, China, IEEE CS Press, USA, 2000.

[4] R. Buyya, D. Abramson, and J. Giddy, *An Economy Driven Resource Management Architecture for Global Computational Power Grids*, Proceedings of the International Conference on Parallel and Distributed Processing Techniques and Applications (PDPTA 2000), June 26-29, 2000, Las Vegas, USA, CSREA Press, USA, 2000.

[5] R. Buyya, J. Giddy, and D. Abramson, *An Evaluation of Economy-based Resource Trading and Scheduling on Computational Power Grids for Parameter Sweep Applications*, Proceedings of the 2nd International Workshop on Active Middleware Services (AMS 2000), Kluwer Academic Press, USA, 2000.

[6] R. Buyya and M. Murshed, *GridSim: A Toolkit for the Modeling and Simulation of Distributed Resource Management and Scheduling for Grid Computing*, The Journal of Concurrency and Computation: Practice and Experience (CCPE), Wiley Press, May 2002.

[7] F. Howell and R. McNab, *SimJava: A Discrete Event Simulation Package For Java With Applications In Computer Systems Modelling*, First International Conference on Web-based Modelling and Simulation, San Diego, CA, Society for Computer Simulation, January 1998.

[8] SPEC, *SPEC CPU2000 Results*, http://www.specbench.org/osg/cpu2000/results/cpu2000.html, Jan. 30, 2002.

[9] R. Buyya, D. Abramson, and J. Giddy, *An Economy Grid Architecture for Service-Oriented Grid Computing*, 10th IEEE International Heterogeneous Computing Workshop (HCW 2001), SF, California, USA, April 2001.

[10] B. Chun and D. Culler, *REXEC: A Decentralized, Secure Remote Execution Environment for Clusters*, Proceedings of 4th Workshop on Communication, Architecture, and Applications for Network-based Parallel Computing, Toulouse, France, January 2000.

[11] B. Chun and D. Culler, *User-centric Performance Analysis of Market-based Cluster Batch Schedulers*, Proceedings of 2nd IEEE International Symposium on Cluster Computing and the Grid, Berlin (CCGrid 2002), Germany, May 2002.

[12] Veridian Systems, OpenPBS v2.3: *The Portable Batch System Software*, Veridian Systems, Inc., Mountain View, CA, September 2000. http://www.openpbs.org/scheduler.html

[13] B. Bode, D. Halstead, R. Kendall, Z. Lei, D. Jackson, *The Portable Batch Scheduler and the Maui Scheduler on Linux Clusters*, Proceedings of the 4th Linux Showcase and Conference, Atlanta, USA, October 2000. http://www.usenix.org/publications/library/proceedings/als2000/full_papers/bode/

[14] George Ma and Paul Lu. PBSWeb:A Web-based Interface to the Portable Batch System, 12th IASTED International Conference on Parallel and Distributed Computing and Systems (PDCS), Las Vegas, Nevada, U.S.A., November 6-9, 2000. http://www.cs.ualberta.ca/~pinchak/PBSWeb/

[15] W. Gentzsch, *Sun Grid Engine (SGE): A Cluster Resource Manager*, http://gridengine.sunsource.net/

[16] Platform, Load Sharing Facility (LSF), http://www.platform.com/products/wm/LSF/

[17] C. Waldspurger, T. Hogg, B. Huberman, J. Kephart, and W. Stornetta, *Spawn: A Distributed Computational Economy*, IEEE Transactions on Software Engineering, Vol. 18, No. 2, IEEE CS Press, USA, February 1992.

[18] M. Stonebraker, R. Devine, M. Kornacker, W. Litwin, A. Pfeffer, A. Sah, and C. Staelin, *An Economic Paradigm for Query Processing and Data Migration in Mariposa*, Proceedings of 3rd International Conference on Parallel and Distributed Information Systems, Austin, TX, USA, 28-30 Sept. 1994, IEEE CS Press, 1994.

[19] B. Chun and D. Culler, *Market-based Proportional Resource Sharing for Clusters*, University of California at Berkeley, Computer Science Division, Technical Report CSD-1092, January 2000.

[20] Libra, The Libra Cluster Scheduler Report and User Manual, http://www.buyya.com/libra/